\pdfoutput=1

\documentclass[entropy,article,accept,moreauthors,pdftex,12pt,a4paper]{mdpi}
\usepackage{graphicx}
\usepackage{amsmath}
\usepackage{bbm}
\usepackage{amsfonts}
\usepackage{amssymb}

\lastpage{22}
\doinum{10.3390/e16010001}
\pubvolume{16}
\pubyear{2014}
\history{This is arXiv postprint, the article is published in Entropy, 2014}
\renewcommand{\emph} [1] { {#1} }

\Title{Entropy and equilibria in competitive systems}

\Author{A. Y. Klimenko$^{*}$}

\address{The University of Queensland, SoMME, QLD 4072, Australia}

\corres{E-mail: klimenko@mech.uq.edu.au}

\abstract{This paper investigates applicability of thermodynamic concepts and principles to competitive systems. We show that Tsallis entropies 
are suitable for characterisation of systems with transitive competition when mutations deviate from Gibbs mutations. Different types of equilibrium
in competitive systems are considered and analysed. As competition rules become more and more intransitive, thermodynamic analogies are eroded and
the behaviour of the system can become complex. This work analyses the phenomenon of punctuated evolution in the context 
of the competitive risk/benefit dilemma.    
}


\begin{document}

%
%
%
%
%
%
%
%
%
%
%
%
%
%
%
%
%
%
%
%
%
%
%
%
%
%
%
%
%
%
%
%
%
%
%
%
%
%
%
%
%
%
%
%
%
%
%
%
%
%
%
%
%
%
%
%
%
%
%
%
%
%
%
%
%
%
%
%
%
%
%
%
%
%
%
%
%

\section{Introduction}

The question of whether systems involving competition can be characterised
\ by quantities resembling conventional thermodynamic parameters does not have
a simple unambiguous answer. This problem was investigated in ref.
\cite{K-PS2012} and it was found that such characterisation is possible under
conditions of transitive competition but, as the system becomes more and more
intransitive, the thermodynamic analogy weakens. The similarity with
conventional thermodynamic principles is strongest when mutations present in
the system belong to the class of Gibbs mutations. While deploying the
conventional logarithmic definition of entropy, the analysis of ref.
\cite{K-PS2012} misses an important point: when mutations deviate from the
Gibbs mutations, the family of Tsallis entropies \cite{Tsallis1} represents a
very convenient choice of entropy to treat these cases. This omission is
rectified in the present work. We also note that Tsallis entropy has been
recently used in modelling of biological replications \cite{KK2013}.

Thermodynamics is strongly linked to the concept of equilibrium. Competitive
systems allow the introduction of different types of equilibrium, possessing
different degrees of similarity with the concept of equilibrium in
conventional thermodynamics. The current work discusses possible cases of
competitive equilibria and performs a detailed analysis based on Tsallis
entropy of the equilibrium through a point of contact, which is more similar
to conventional thermodynamics than the other cases.

From the thermodynamic perspective, the present work is only an example of
using Tsallis entropy. We do not attempt to draw any general thermodynamic
conclusions and the use of non-extensive entropy in other applications may
well be different from our treatment of equilibria in completive systems. The
problem of general\ consistency between physical equilibrium conditions and
definitions of non-extensive entropies has been analysed by Abe
\cite{Abe2,Abe3}. Non-extensive statistical mechanics has been reviewed by
Tsallis \cite{Tsallis1}, while non-extensive entropies associated with this
mechanics are discussed in refs.
\cite{Tsallis1,KK2013,Abe2,Abe3,Tsallis2,Thurner1} and many other publications.

The last section deals with intransitive cases when the thermodynamic analogy
weakens and the possibility of using entropy as a quantity that always tends
to increase in time or remain constant is not assured. This section analyses
the \textit{risk/benefit dilemma} represented by a competitive system, whose
evolution can be transitive or intransitive depending on the choice of the
system parameters. In the intransitive case, the evolution of the system
appears to be punctuated by sudden collapses and becomes cyclic. This
\textit{punctuated evolution} is similar to the concept of \textit{punctuated
equilibrium} in evolutionary biology \cite{PE1993Nature}, although in the
context of thermodynamics the latter term might be misleading as the system is
not in equilibrium and keeps evolving between the punctuations.

\section{Competitive systems}

Competitive systems involve the process of competition in its most generic
form. The elements of competitive systems compete with each other according to
preset rules. The rules define the winners and losers for each competition
round based on properties of the elements denoted here by $\mathbf{y}$. The
properties of the losers are lost while the winners utilise the resource
vacated by the losers and duplicate their properties. The process of
duplication is not perfect and involves random mutations, which are mostly
detrimental for competitiveness of the elements. The expression A$\,\prec\,$B
(or equivalently $\mathbf{y}_{\text{A}}\prec\mathbf{y}_{\text{B}}$) indicates
that element B with properties $\mathbf{y}_{\text{B}}$ is the winner in
competition with element A with properties $\mathbf{y}_{\text{A}}$. If two
elements have equivalent strength $\mathbf{y}_{\text{A}}\simeq\mathbf{y}%
_{\text{B}}$, the winner is to be determined at random. In computer
simulations, the elements are also called Pope particles and exchange of
properties is called mixing by analogy with the conventions adopted in
particle simulations of reacting flows. Two forms of mixing --- conservative
and competitive --- can be distinguished. The former is predominantly used in
the flow simulations while the latter is associated with competitive systems.
The rest of this section introduces basic terms used in characterisation of
competitive systems; further details can be found in refs.
\cite{K-PS2012,K-PT2013}.

\begin{figure}[h]
\begin{center}
\includegraphics[width=14cm,page=1,trim=5.5cm 6cm 6cm 4cm, clip ]{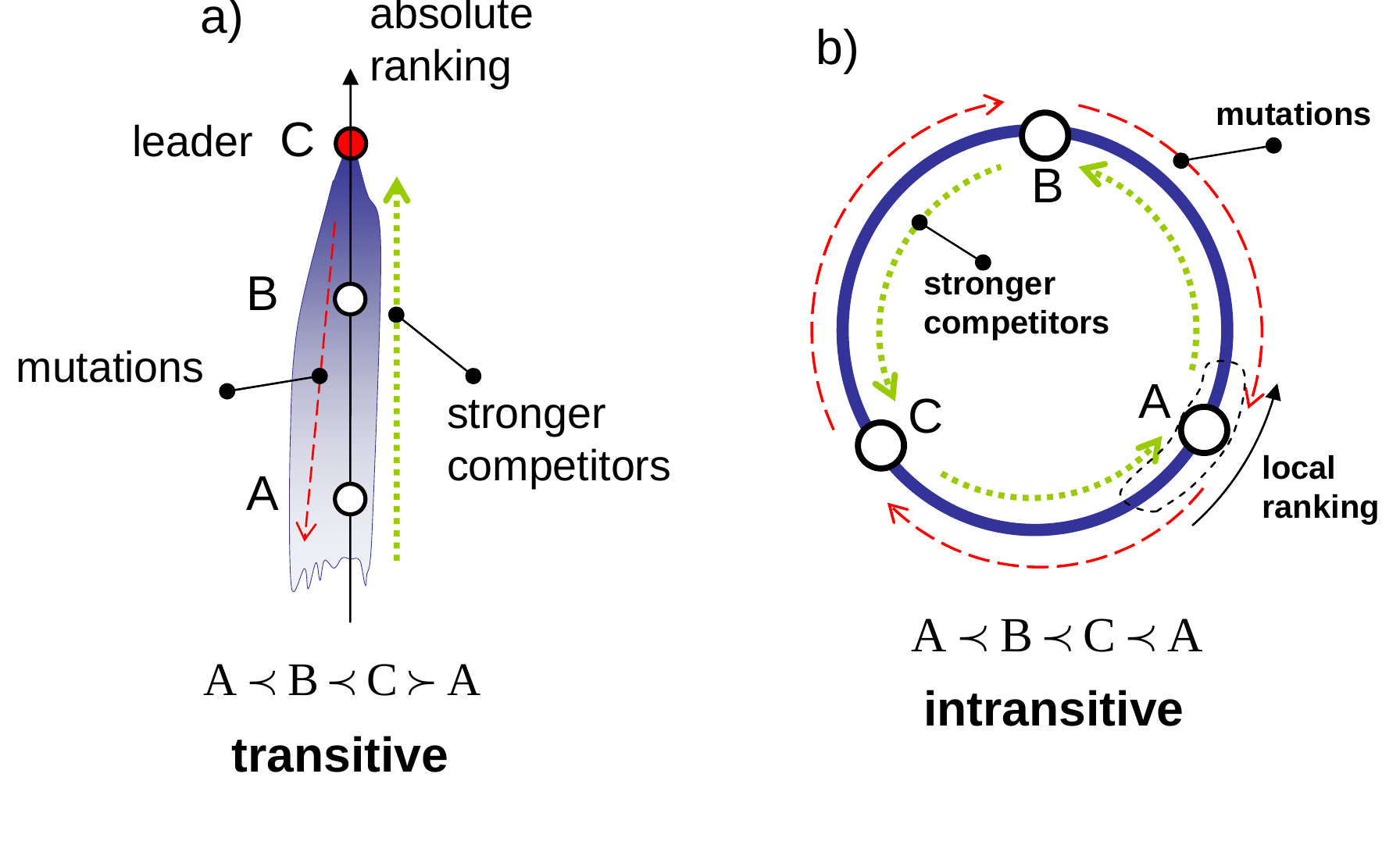}
\caption{Examples of systems with a) transitive and b) intransitive competitions.  }
\label{fig1}
\end{center}
\end{figure}

The competition rules are divided into two major categories: transitive and
intransitive. In transitive competitions superiority of B over A and C over B
inevitably demands superiority of C over A, that is
\begin{equation}
\mathbf{y}_{\text{A}}\preceq\mathbf{y}_{\text{B}}\preceq\mathbf{y}_{\text{C}%
}\Longrightarrow\mathbf{y}_{\text{A}}\preceq\mathbf{y}_{\text{C}}%
\end{equation}
As illustrated in figure \ref{fig1}a, transitive competitions enable
introduction of an absolute ranking $r(\mathbf{y}),$ which is a numerical
function that determines superior (stronger) and inferior (weaker) elements:
\begin{equation}
\mathbf{y}_{\text{A}}\preceq\mathbf{y}_{\text{B}}\text{ }\Longleftrightarrow
r(\mathbf{y}_{\text{A}})\leq r(\mathbf{y}_{\text{B}})
\end{equation}
The competitive transformations can be interpreted as reactions between the
particles
\begin{equation}
\text{A}+\text{B}\longrightarrow\text{B}^{\prime}+\text{B},\;\;\;\;\text{A}%
\prec\text{B}%
\end{equation}
where B$^{\prime}$ is different from B due to mutations. B is the winner in
competition with A and, thus, B is entitled to occupy the resource (i.e.
particle) previously occupied by A. Properties of A are lost and the
properties of B are copied across into A. (Conservative properties, which are
not considered here, would be transferred in the opposite direction from the
loser to the winner.) The copying process is not perfect due to mutations,
which are random alterations of properties of B during copying. If mutations
are not present then B$^{\prime}$=B. Unlike random walks, mutations have a
strong preference for the negative directions: it is likely (in case of
non-positive mutations, it is certain) that B$^{\prime}\,\preceq\,$B. If rare
positive mutations B$^{\prime}\,\succ\,$B are present, the distribution of
particles may escalate towards higher ranks when the leading particle (i.e the
particle with the maximal absolute ranking in the group) is occasionally
overtaken by a new leader. One of the main results of ref. \cite{K-PS2012} is
linking absolute ranking to the entropy potential $s_{y}=s_{y}(r)$ and, under
some restrictions (e.g. Gibbs mutations), proving the associated competitive H-theorem.

The competition rules, however, are not necessarily transitive and the
competition is deemed intransitive when at least one intransitive triplet
\begin{equation}
\mathbf{y}_{\text{A}}\preceq\mathbf{y}_{\text{B}}\preceq\mathbf{y}_{\text{C}%
}\prec\mathbf{y}_{\text{A}} \label{int3}%
\end{equation}
exists (see figure \ref{fig1}b). Although intransitive competitions do not
generally permit absolute ranking of elements, they can be characterised by a
coranking function
\begin{equation}
\mathbf{y}_{\text{A}}\preceq\mathbf{y}_{\text{B}}\text{ }\Longleftrightarrow
\rho(\mathbf{y}_{\text{A}},\mathbf{y}_{\text{B}})\leq0
\end{equation}
which, by definition, should be antisymmetric $\rho(\mathbf{y}_{\text{A}%
},\mathbf{y}_{\text{B}})=-\rho(\mathbf{y}_{\text{B}},\mathbf{y}_{\text{A}})$.
In the case of transitive competitions, the coranking function can be
expressed in terms of absolute ranking by
\begin{equation}
\rho(\mathbf{y}_{\text{A}},\mathbf{y}_{\text{B}})=r(\mathbf{y}_{\text{A}%
})-r(\mathbf{y}_{\text{B}})
\end{equation}
In addition to $\rho(\mathbf{y}_{\text{A}},\mathbf{y}_{\text{B}}),$ it is
useful to define sharp coranking
\begin{equation}
R(\mathbf{y}_{\text{A}},\mathbf{y}_{\text{B}})=\operatorname{sign}\left(
R(\mathbf{y}_{\text{A}},\mathbf{y}_{\text{B}})\right)  =\left\{
\begin{array}
[c]{c}%
-1\;\;\text{if }\mathbf{y}_{\text{A}}\prec,\mathbf{y}_{\text{B}}\\
0\;\;\text{if }\mathbf{y}_{\text{A}}\simeq\mathbf{y}_{\text{B}}\\
+1\;\;\text{if }\mathbf{y}_{\text{A}}\succ\mathbf{y}_{\text{B}}%
\end{array}
\right.
\end{equation}
In case of two-particle mixing, evolution of the system is determined by the
sharp coranking $R$. The graded coranking can be useful in establishing
relative ranks within each mixing group, when mixing of multiple particles is considered.

The distributions of elements in the property space is characterised by the
particle distribution function $\varphi(\mathbf{y})=nf(\mathbf{y})$ where $n$
is the total number of particles and $f(\mathbf{y})$, which can be interpreted
as the probability distribution function (pdf), satisfies the normalisation
condition
\begin{equation}
\int\limits_{\infty}f(\mathbf{y})d\mathbf{y}=1
\end{equation}
When a competitive system is divided into $K$ subsystems $I=1,2,...,K$ and
each subsystem $I$ has the $b_{I}$-th fraction of the particles, we may
characterise each of these subsystems by its own normalised distribution
$\phi_{I}(\mathbf{y});$ that is
\begin{equation}
f(\mathbf{y})=\sum_{I=1}^{K}b_{I}\phi_{I}(\mathbf{y}),\;\;\varphi
(\mathbf{y})=nf(\mathbf{y})=\sum_{I=1}^{K}\varphi_{I}(\mathbf{y}),\;
\label{grp1}%
\end{equation}%
\begin{equation}
\ \varphi_{I}(\mathbf{y})=n_{I}\phi_{I}(\mathbf{y}),\;b_{I}=\frac{n_{I}}%
{n},\;\;\int\limits_{\frak{D}_{I}}\phi_{I}(\mathbf{y})d\mathbf{y}=1
\label{grp2}%
\end{equation}
The subsystems can be distinguished by having different domains $\frak{D}_{I}$
or by other means. When subsystems are distinguished, it is useful to define
the coranking of the distributions
\begin{equation}
\bar{R}([\phi_{I}],[\phi_{J}])=\underset{\infty}{\int\int}R(\mathbf{y}%
,\mathbf{y}^{\prime})\phi_{I}(\mathbf{y})\phi_{J}(\mathbf{y}^{\prime
})d\mathbf{y}d\mathbf{y}^{\prime}%
\end{equation}
which indicate relative strength of subsystem distributions with respect to
each other. We can say ''the subsystem $I$ is stronger than the subsystem
$J$'' and write $[\phi_{I}]\succ\lbrack\phi_{J}]$ when $\bar{R}([\phi
_{I}],[\phi_{J}])>0$. Note that the subsystem coranking is antisymmetric
$\bar{R}([\phi_{I}],[\phi_{J}])=-\bar{R}([\phi_{J}],[\phi_{I}])$ and
self-neutral $\bar{R}([\phi_{I}],[\phi_{I}])=0.$

Examples of systems using competitive mixing can be found in refs.
\cite{K-PS2010,KP2012,K-PS2012, K-PT2013}

\section{Competition and $q$-exponential distributions}

We consider transitive competition with elements possessing a scalar property
$y,$ which is selected to be aligned with ranking (that is $r(y)$ is a
monotonically increasing function and the absolute ranking is effectively
specified by $y$). Hence, for any two elements A and B
\begin{equation}
\text{A}\preceq\text{B }\Longleftrightarrow y_{\text{A}}\leq y_{\text{B}}%
\end{equation}
The problem is deemed to be uniform with respect to shifts along $y$. Assuming
that mutations, which are originated at point $y^{\prime}$ and distributed
with the probability density function $f_{m}(y,y^{\prime}),$ are uniform
$f_{m}(y,y^{\prime})=f_{m}(y-y^{\prime}),$ the general competitive evolution
equation \cite{K-PS2012} takes a more simple form given by%
\begin{equation}
\frac{\partial f(y)}{\partial t}=\int_{y}^{0}f_{m}(y-y^{\prime})F(y^{\prime
})f(y^{\prime})dy^{\prime}-\left(  1-F(y)\right)  f(y) \label{GE1}%
\end{equation}
The competitive evolution equation specifies a balance between mutations,
given by the first term on the right hand side of this equation, and the
losses due to competition, given by the second term. The function $F$ is the
cdf (cumulative distribution function) of the pdf $f$%
\begin{equation}
F(y)=\int_{-\infty}^{y}f(y^{\prime})dy^{\prime} \label{cdf}%
\end{equation}
If mutations are non-positive then
\begin{equation}%
\begin{array}
[c]{cc}%
f_{m}(y)\geq0 & \text{if }y\leq0\\
f_{m}(y)=0 & \text{if }y>0
\end{array}
\label{posM}%
\end{equation}
Equation (\ref{GE1}) can be integrated to yield
\begin{align}
\frac{\partial F(y)}{\partial t}  &  =\frac{1}{2}\int_{-\infty}^{0}%
F_{m}(y-y^{\prime})\frac{\partial F^{2}(y^{\prime})}{\partial y^{\prime}%
}dy^{\prime}-F(y)+\frac{F^{2}(y)}{2}\nonumber\\
&  =\frac{F_{m}(y)+F^{2}(y)}{2}-F(y)+\frac{1}{2}\int_{y}^{0}f_{m}(y-y^{\prime
})F^{2}(y^{\prime})dy^{\prime} \label{GE2}%
\end{align}
where $F_{m}$ is the cdf that corresponds to pdf $f_{m}$.

The Gibbs mutations \cite{K-PS2012} correspond to $q=1$ implying that the
distribution $f_{m}(y)$ is based in this case on the conventional exponent
\begin{equation}
f_{m}(y)=\exp(y)H\left(  -y\right)
\end{equation}
where
\begin{equation}
H(y)=\left\{
\begin{array}
[c]{cc}%
1 & \text{if }y\geq0\\
0 & \text{if }y<0
\end{array}
\right.
\end{equation}
is the Heaviside function. Note that there is no loss of generality in setting
$\alpha=1$ in $f_{m}\sim\exp(\alpha y)$ since the variable $y$ can always be
rescaled to eliminate $\alpha$. In case of Gibbs mutations, the pdf $f$ is
given by
\begin{equation}
f(y,y^{\ast})=\exp(y-y^{\ast})H(y^{\ast}-y)
\end{equation}
where $y^{\ast}$ is the position of leading particle.

\begin{figure}[h]
\begin{center}
\includegraphics[width=7.5cm]{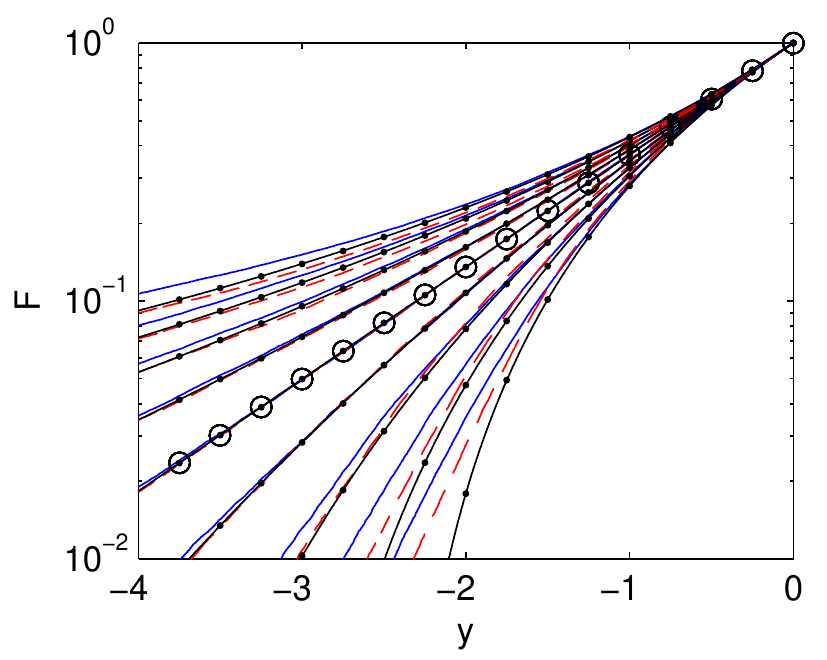}
\caption{Simulated long- and short-tailed distributions in comparison with q-exponents: 
solid curves --- simulated for mutations I, dased curves --- simulated for mutations II, solid curves with dots --- q-exponents. 
The cdf are plotted for the values of $Q = \{0.6, 0.7, 0.8, 0.9, 1, 1.1, 1.2, 1.3, 1.4 \}$; the curves are numbered from the bottom to the top.
The circles mark the line corresponding to $Q=1$.}
\label{figm1}
\end{center}
\end{figure}

\begin{figure}[h]
\begin{center}
\includegraphics[width=14cm,page=2,trim=3cm 5cm 4cm 3cm, clip]{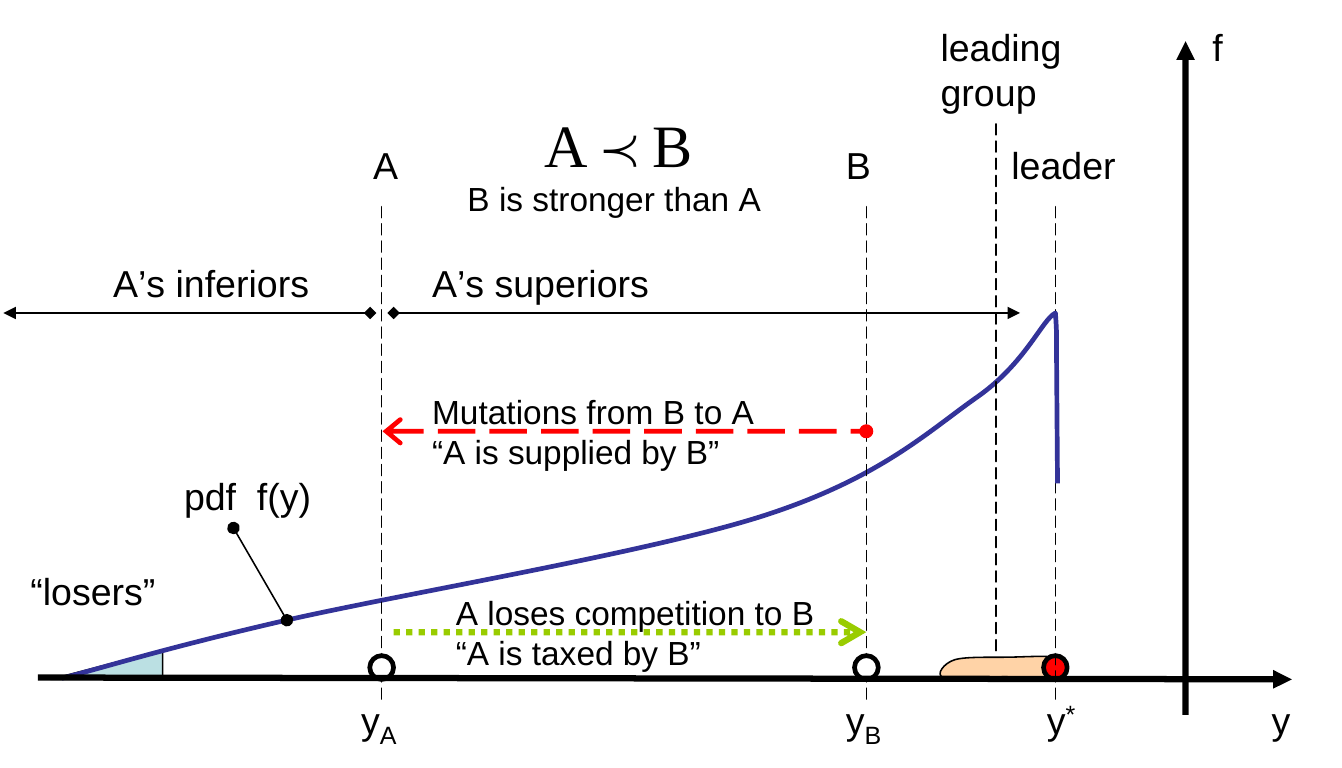}
\caption{Distribution of the elements and exchanges between them in a transitive competition.}
\label{fig2}
\end{center}
\end{figure}

In this work we are interested in the case when the pdf $f$ can be
approximated by the $q$-exponential distribution
\begin{equation}
f(y,y^{\ast})=f_{q}(y-y^{\ast})=\exp_{q}\left(  \frac{y-y^{\ast}}{2-q}\right)
H(y^{\ast}-y) \label{fq}%
\end{equation}
where
\[
e_{q}^{y}=\exp_{q}(y)=\left(  1+\left(  1-q\right)  y\right)  ^{\frac{1}{1-q}}%
\]
is the so called $q$-exponent and
\[
\ln_{q}(y)=\frac{y^{1-q}-1}{1-q}%
\]
is the corresponding $q$-logarithm. If $q\rightarrow1$ then the $q$-functions
approach the conventional $\exp(y)$ and $\ln(y)$. The cdf, corresponding to
pdf (\ref{fq}) is given by
\begin{equation}
F(y,y^{\ast})=F_{Q}(y-y^{\ast})=\left\{
\begin{array}
[c]{cc}%
\exp_{Q}(y-y^{\ast}) & \text{if }y\leq y^{\ast}\\
1 & \text{if }y>y^{\ast}%
\end{array}
\right.  \label{FQ}%
\end{equation}
where $Q=1/(2-q)$. The distribution (\ref{fq}) solves the governing equation
(\ref{GE1}) with an asymptotic precision of $O((1-q)^{2})$ provided the
mutations are distributed according to
\begin{equation}
f_{m}(y)=\left(  \left(  3-2q\right)  e_{q}^{y}-2\left(  1-q\right)
e_{q}^{2y}\right)  H(-y)
\end{equation}%
\begin{equation}
F_{m}(y)=\left\{
\begin{array}
[c]{cc}%
\left(  \left(  2-Q\right)  e_{Q}^{y/Q}-\left(  1-Q\right)  e_{Q}%
^{2y/Q}\right)  & \text{if }y\leq0\\
1 & \text{if }y\geq0
\end{array}
\right.
\end{equation}
Figure \ref{figm1} illustrates that, as expected, the cdf of simulated
distributions are very close the corresponding $q$-exponents when $q$ is close
to unity. The $q$-exponential functions can also serve as very good
approximations for distribution in competitive systems for a wide range of
$q$. Consider the $q$-exponential distribution of mutations\
\begin{equation}
f_{m}(y)=\exp_{q^{\prime}}\left(  \frac{y}{2-q^{\prime}}\right)  H(-y)
\end{equation}
with the cdf $F_{m}(y),$ which is given by $q$-exponential functions similar
to (\ref{FQ}) with $Q^{\prime}=1/(2-q^{\prime})$ The approximate solutions
shown in figure \ref{figm1} correspond to $q$-exponential (\ref{fq}) with
\begin{equation}
q=\frac{2+q^{\prime}}{4-q^{\prime}},\;\;Q-1=\frac{2}{3}\left(  Q^{\prime
}-1\right)  \label{qqm}%
\end{equation}
The cdf shapes presented in Figure \ref{figm1} indicate that, although
$q$-exponential distributions are not necessarily exact for competitive
systems, they are reasonably accurate and correspond very well to the physical
nature of the problem when mutations deviate from Gibbs mutations.\ In the
competitive system illustrated in figure \ref{fig2}, every location is taxed
due to competition with superior elements and at the same time is supplied by
mutations originated at superior elements. For Gibbs mutations, the
competitive system schematically depicted in figure \ref{fig2} is in the state
of detailed balance: every location is taxed and supplied at the same rate by
any given superior. In simple systems with constant a priori phase space
$A(y)$, the Gibbs mutations are distributed exponentially ($q=1$). When
mutations deviate from Gibbs mutations, the overall rates of taxing and
supplying must negate each other under steady conditions but there is no
detailed balance in relations with different groups of superiors. For
long-tailed (superexponential) distributions with $q>1$, weak particles are
supplied more by the leaders and are taxed more by immediate superiors. For
short-tailed (superexponential) distributions with $q<1$, weak particles are
supplied more by the immediate superiors and are taxed more by the leaders.

\begin{figure}[h]
\begin{center}
\includegraphics[width=6cm]{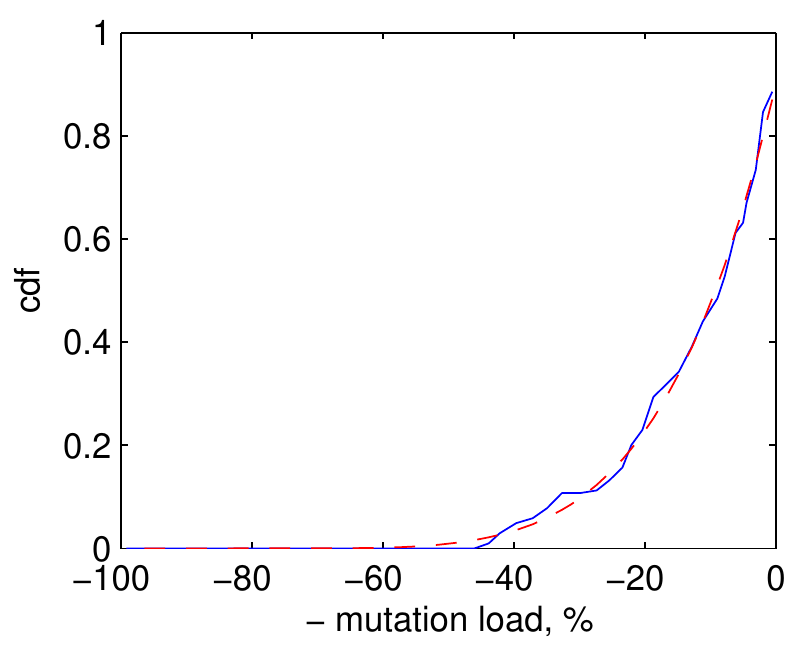}
\caption{The cdf of the experimental \cite{mitdist2008} distribution of the A3243G
mtDNA mutation (solid line) compared to the q-exponential cdf $\exp_Q{(k y /Q)}$ with $Q=0.8$ and $k/Q=6$  (dashed line). 
Since these mutations are deleterious, their extent is shown as negative.}
\label{genet}
\end{center}
\end{figure}

Competitive systems are aimed at studying generic properties of systems with
competition and mutations. Although we do not specifically intend to model
distributions of biological mutations, these distributions are still of some
interest here as real--world examples of complex competitive systems. Ohta
\cite{Ohta1977} considered near-neutral genetic mutations and suggested that
these mutations have exponential distributions. Modern works tend to use the
Kimura distribution\cite{mitdist2008}, which has a complicated mathematical
form, deviates from pure exponents and, theoretically, corresponds to a
genetic drift of neutral mutations. It seems that the reported distributions
of genetic mutations tend to be slightly subexponential. Figure \ref{genet}%
\ illustrates that the experimental distribution of mutation A3243G of
mitochondrial DNA in humans \cite{mitdist2008} is well approximated by
$q$-exponential cdf with $Q=0.8.$ Since these mutations are known to be
deleterious \cite{mitmut2009}, they are shown as negative in the figure (in
agreement with the notations adopted in the rest of the present work).

\section{Tsallis entropy in competitive systems}

\emph{Free} Tsallis entropy in competitive systems is defined by
\begin{equation}
S([f])=\;\int_{\infty}\left(  \tilde{f}(\mathbf{y})\ln_{q}\left(  \frac
{1}{\tilde{f}(\mathbf{y})}\right)  +\tilde{f}(\mathbf{y})^{\gamma}%
s_{y}(\mathbf{y})\right)  A(\mathbf{y})d\mathbf{y}\; \label{Scs}%
\end{equation}
with two likely choices of the exponent $\gamma$ given by $\gamma=1$ and
$\gamma=q$. Here we denote $\tilde{f}(\mathbf{y})=f(\mathbf{y})/A(\mathbf{y}%
).$\ The first term in the integral is the configurational entropy, which
represents the randomising influence of mutations, while the second term
involves the entropy potential $s_{y}(\mathbf{y}),$ which reflects the
influence of competition (i.e. $s_{y}$ increases with $r$ reflecting higher
likelihood of survival of more competitive elements). The term $\tilde
{f}(\mathbf{y})^{\gamma}s_{y}(\mathbf{y})$ in (\ref{Scs}), which can be called
the escort term, reflects nature's preference for elements with higher
rankings (for example, in biological systems, ranking reflects fitness).
\emph{Due to presence of the potential $s_{y}(y)$, the competitive entropy
defined by (\ref{Scs}) is analogous to free entropy of conventional
thermodynamics, which is proportional to free energy (Gibbs or Helmholtz)
taken with the negative sign. The distribution and entropy that correspond to
$s_{y}=0$ (i.e. not affected by competition) can be termed ''a priori''
bearing some resemblance to prior probabilities in Bayesian inference. }The
definition of entropy in (\ref{Scs}) is of Boltzmann type, i.e. implying
validity of the \textit{Stosszahlansatz} (stochastic independence of particles
from each other). Variation of the distribution function results in
\begin{equation}
\chi_{y}=\frac{\delta S}{\delta f(\mathbf{y})}=\frac{1}{A(\mathbf{y})}%
\frac{\delta S}{\delta\tilde{f}(\mathbf{y})}=\ln_{q}\left(  \frac{1}%
{c_{q}\tilde{f}(\mathbf{y})}\right)  +\gamma\tilde{f}(\mathbf{y})^{\gamma
-1}s_{y}(\mathbf{y}) \label{VarS}%
\end{equation}
where
\begin{equation}
c_{q}=q^{\frac{1}{q-1}}\rightarrow e\text{ as }q\rightarrow1
\end{equation}
and $\chi_{y}$ can be interpreted as the competitive potential of state $y$.
Maximisation of $S$ constrained by the normalisation
\begin{equation}
Z_{f}([f])=\int_{\infty}\tilde{f}(\mathbf{y})A(\mathbf{y})dy=1
\end{equation}
and by the location of the leading element, that is
\begin{equation}
f(\mathbf{y})=0\text{ \ for any }\mathbf{y}\succ\mathbf{y}^{\ast}
\label{flead}%
\end{equation}
results in the following condition
\begin{equation}
\frac{\delta S}{\delta\tilde{f}(\mathbf{y})}+\lambda\frac{\delta Z_{f}}%
{\delta\tilde{f}(\mathbf{y})}=0 \label{VarSL}%
\end{equation}
where $\lambda$ is the Lagrange multiplier, implying that the local
competitive potential
\[
\chi_{y}=\frac{1}{A(\mathbf{y})}\frac{\delta S}{\delta\tilde{f}(\mathbf{y}%
)}=-\frac{\lambda}{A(\mathbf{y})}\frac{\delta Z_{f}}{\delta\tilde
{f}(\mathbf{y})}=-\lambda
\]
is the same everywhere in equilibrium. Consider a simple case of scalar $y$
and constant a priori capacity $A=\operatorname{const}.$

\begin{enumerate}
\item \textbf{The translational case of} $\gamma=1.$ With (\ref{VarS}),
equation (\ref{VarSL}) takes the form
\[
\ln_{q}\left(  \frac{1}{c_{q}\tilde{f}(y)}\right)  +s_{y}(y)+\lambda=0
\]
that with $s_{y}(y)=ky,\,\ $ $A=c_{q}q$ and $\lambda=-ky^{\ast}$ results in
the pdf and cdf given by
\begin{equation}
f(y,y^{\ast})=\frac{qkH(y^{\ast}-y)}{\exp_{q}(k\left(  y^{\ast}-y\right)
)}=qkH(y^{\ast}-y)\exp_{q_{2}}\left(  k\left(  y-y^{\ast}\right)  \right)
\end{equation}%
\begin{equation}
F(y,y^{\ast})=\left\{
\begin{array}
[c]{cc}%
\exp_{Q}(k\left(  y-y^{\ast}\right)  /Q) & \text{if }y\leq y^{\ast}\\
1 & \text{if }y>y^{\ast}%
\end{array}
\right.  \label{FQk}%
\end{equation}
where $Q=1/q$ and $q_{2}=2-q$. Since $y^{\ast}$ is arbitrary in this case, the
distribution can be freely shifted along $y.$ The location of $y^{\ast}$ is
determined by (\ref{flead}).

\item \textbf{The multiplicative case of} $\gamma=q.$ Equations (\ref{VarS})
and (\ref{VarSL}) take the form
\begin{equation}
\ln_{q}\left(  \frac{\exp_{q}(s_{y}(y))}{c_{q}\tilde{f}(y)}\right)  +\lambda=0
\end{equation}
that, with, $A=1$ and $s_{y}(y)=k\left(  y-y^{\ast}\right)  $, results in the
pdf
\begin{equation}
f(y,y^{\ast})=\frac{H(y^{\ast}-y)}{Z_{q}}\exp_{q}(k\left(  y-y^{\ast}\right)
)
\end{equation}
with arbitrary value of $Z_{q}$ depending on $\lambda$. This value can be
determined from the normalisation requiring that $Z_{q}=k^{-1}/(2-q).$ The
corresponding cdf $F_{Q}(y,y^{\ast})$ is the same as (\ref{FQk}) with the
$q$-parameter given by $Q=Z_{q}k=1/(2-q).$
\end{enumerate}

In case of physical thermodynamics, Tsallis et. al. \cite{Tsallis2} recommend
using $\gamma=q$ in conjunction with the escort distribution for \emph{the}
energy constraints as the best option. Competitive thermodynamics\emph{,} as
considered here\emph{,} does not have any energy constraints (assuming that
the conservative properties are limited to the number of particles, we do not
have any energy defined for the system) and selection of $\gamma$ needs to be
considered again. The choice of $\gamma$ for competitive systems is determined
by the physics of the problem and can be different for different processes. If
infrequently positive mutations are present and the distribution with fixed
number of particles escalates by gradually increasing $y^{\ast}$ in time, then
$\gamma=1$ is preferable. Indeed, while $y^{\ast}$ increases, the definition
of entropy remains exactly the same and the escalation is seen as a natural
process of increasing entropy in the system. If $\gamma=q$, the definition of
entropy is dependent on the position of the leading particle. The choice of
\ $\gamma=q$ is more suitable for competitions between subsystems placed at
fixed locations but with the numbers of particles that can be altered due to
exchanges. Gibbs mutations correspond to $q=1$ and the choices $\gamma=1$ and
$\gamma=q$ coincide in this case. In the previous work \cite{K-PS2012}, the
Boltzmann-Gibbs entropy was used for non-Gibbs mutations by artificially
making \emph{the} phase volume dependent on the leading particle position
$A=A(\mathbf{y},\mathbf{y}^{\ast})$. Unlike the Tsallis entropy considered in
the present work, the old treatment of the problem \cite{K-PS2012} did not
allow for a unified definition of entropy valid for different $\mathbf{y}%
^{\ast}$ (i.e. the Boltzmann-Gibbs entropy provides a unified, $\mathbf{y}%
^{\ast}$-independent definition of competitive entropy only for the Gibbs mutations).

\section{Equilibria in competing systems. \label{sec_eql}}

\begin{figure}[h]
\begin{center}
\includegraphics[width=10cm,page=3,trim=3cm 2cm 9cm 2cm, clip]{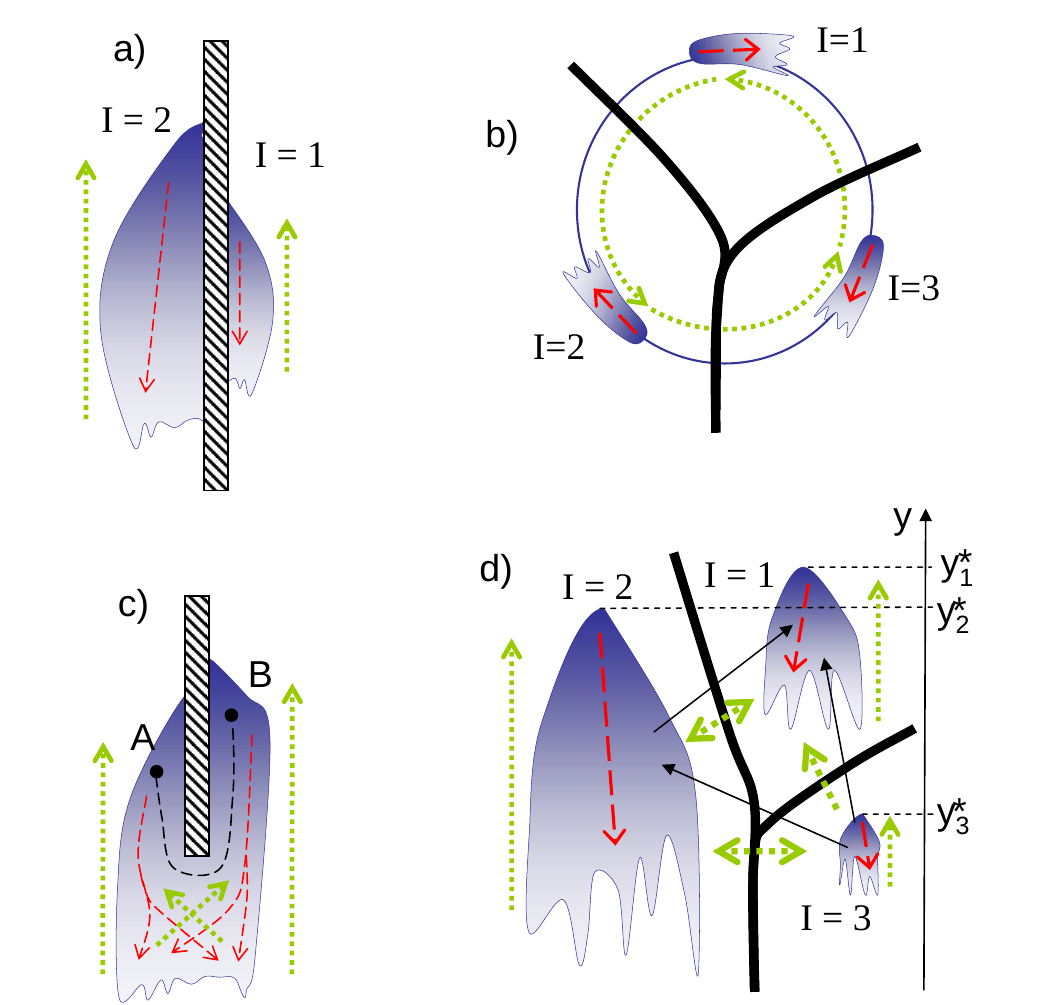}
\caption{Equilibrium in competitive systems: a) isolated, b) competing and c) connected, 
while case d) illustrates the impossibility of 
competing equilibrium in transitive competition. Dashed arrows show the direction of mutations, 
dotted arrows indicate transport of particles due to competition.}
\label{fig3}
\end{center}
\end{figure}

A competitive system can be divided into subsystems and the question of
equilibrium conditions between these subsystems appears. If the system is
subdivided into $K$ subsystems $I=1,2,...,K$ and subsystem $I$ has the $b_{I}%
$-th fraction of the particles, we may characterise each of these subsystems
by its own normalised distribution $\phi_{I}(\mathbf{y})$ as specified by
equations (\ref{grp1}) and (\ref{grp2}).\ Assuming that equilibrium or
steady-state conditions are achieved within each subsystem, the major
equilibrium cases include:

\begin{enumerate}
\item \textbf{Equilibria in isolated subsystems.} Isolated subsystems do not
exchange mutations and do not compete against each other. Equilibria are
established in isolated subsystems independently of the other subsystems (see
figure \ref{fig3}a)

\item \textbf{Competing equilibria.} Particles in these subsystems compete
against each other but mutations do not cross the subsystem boundaries as
illustrated in figure \ref{fig3}b. Competing equilibria tend to be less stable
than connected equilibria considered below and, generally, are impossible in
transitive competitions. Indeed, if $y_{I}^{\ast}>y_{J}^{\ast}$ (i.e. the
leading element of subsystem $I$ is more competitive than the leading element
of subsystem $J$), $y_{I}^{\ast}$ cannot lose to any element of subsystem $J$
while $y_{J}^{\ast}$ will eventually lose to leading elements of $I$. There is
no equilibrium in transitive competition depicted in figure \ref{fig3}d since
the subsystem $I=1$ is going to win all particle resources for itself. If
$y_{I}^{\ast}=y_{J}^{\ast}$ then the two leaders will eventually meet in
competition and due to their equivalent strength, the winner of this round
(which ultimately belongs to the winning subsystem) is to be selected at
random. Competing equilibria are nevertheless possible in intransitive
competitions. This, obviously, requires that
\begin{equation}
R_{I}=R([\phi_{I}],[f])=0 \label{eq1}%
\end{equation}
for all $I=1,...,K$ or, otherwise, $n_{I}$ would grow for $R_{I}>0$ and
decrease for $R_{I}<0$. As discussed in Appendix of ref. \cite{K-PS2012},
oscillations are to appear in competing equilibria between subsystems
$1,...,K$, unless
\begin{equation}
R_{IJ}=R([\phi_{I}],[\phi_{J}])=0 \label{eq2}%
\end{equation}
for every $I$ and $J,$ Constraint (\ref{eq2}) implies all subsystems should
have the same relative strength $[\phi_{I}]\simeq\lbrack\phi_{J}],$ which is a
stronger condition than $[\phi_{I}]\simeq\lbrack f]$ required by (\ref{eq1}).
Condition (\ref{eq2}) is necessary to avoid oscillations between the
subsystems. If present, the oscillations can be stable, neutral or unstable.
Example in ref. \cite{K-PS2012} demonstrates the case when oscillations are
unstable. Competing equilibria exist only in competitive systems and, it
seems, do not have an analog in conventional thermodynamics.

\item \textbf{Connected equilibria.} In this case the subsystems $I=1,...,K$
are connected by both competition and mutations. For Gibbs mutations ($q=1$),
the competitive $H$-theorem applies ensuring the detailed balance of the
equilibrium state \cite{K-PS2012}. This implies that, in equilibrium, the
connection between any two elements or groups of elements can be severed
without any effect on the state of the system. Severing connection between two
elements terminates both competition and mutations between these elements.
This is illustrated in figure \ref{fig3}c, where the direct connection between
points A and B is severed, although A and B remain connected through other
elements as shown by the dashed line. The equilibrium conditions are given by
the equivalence of all competitive potentials $\chi_{I}=\chi_{J}$ for every
$I$ and $J$, where the formula for competitive potential
\cite{K-PS2012,K-PT2013}%
\begin{equation}
\chi_{I}=\frac{\partial S}{\partial n_{I}}=\ln\left(  \frac{Z_{I}}{en_{I}%
}\right)  \label{eq3}%
\end{equation}
is obtained by differentiating entropy with respect to $n_{I}$. The partition
functions $Z_{I}$ are evaluated for each subsystem $I$ as integral over the
subsystem domain $\frak{D}_{I}$%
\begin{equation}
Z_{I}=\int_{\frak{D}_{I}}A_{I}(\mathbf{y})\exp\left(  s_{y}(\mathbf{y}%
)\right)  d\mathbf{y} \label{eq3Z}%
\end{equation}
Equilibrium in competitive systems with Gibbs mutations resembles most the
equilibria of conventional thermodynamics. In the case of general non-positive
mutations (i.e. non-Gibbs mutations), the state of the system depends on the
type of contact. Here, we distinguish two cases of interest:

\begin{enumerate}
\item \textbf{Point of contact.} Two subsystems $I$ and $J$ have a point of
contact at $\mathbf{y}=\mathbf{y}^{\circ}$ when the elements from the vicinity
of $\mathbf{y}=\mathbf{y}^{\circ}$ effectively belong to the both subsystems,
while the other elements are isolated within their subsystems. Hence, at
equilibrium the density of particles representing competing elements must be
the same in both subsystems at the point of contact
\begin{equation}
\frac{n_{I}\phi_{I}(\mathbf{y}^{\circ})}{A_{I}(\mathbf{y}^{\circ})}%
=\frac{n_{J}\phi_{J}(\mathbf{y}^{\circ})}{A_{J}(\mathbf{y}^{\circ})}
\label{eq-1}%
\end{equation}
The phase volume associated with the distributions is likely to be the same on
both sides $A_{I}(\mathbf{y}^{\circ})=A_{J}(\mathbf{y}^{\circ})$. Existence of
a single point of contact (or several points of contact that, as discussed
below, do not form a loop while connecting several subsystems) changes $n_{I}$
but does not affect the distributions $\phi_{I}(\mathbf{y})$. More than one
point of contact between two systems with non-Gibbs mutations is likely to
change not only $n_{I}$ but also the distributions $\phi_{I}(\mathbf{y}).$

\item \textbf{Complete merger.} The subsystems are merged into a single system
with the overall stationary distribution $f=f_{0}(\mathbf{y}).$ Unless
mutations are limited to Gibbs mutations, the subsystems are likely to undergo
complex adjustments changing their distributions. \ If the term equilibrium is
used for this steady state, it should be remembered that, generally, there is
no detailed balance in the system. \ The overall stationary distribution is
inseparable: $f_{0}(\mathbf{y})$ may change if the contact between any two
locations is severed. Note that, although unusual, inseparable systems exist
in conventional thermodynamics: objects with negative heat capacity
\cite{K-OTJ2012} may serve as an example.
\end{enumerate}
\end{enumerate}

Among different types of equilibrium in competitive systems, the equilibrium
at a point of contact is most suitable for thermodynamic analysis, even when
mutations substantially deviate from Gibbs mutations.

\section{Entropy for equilibrium through a point of contact}

\begin{figure}[h]
\begin{center}
\includegraphics[width=12cm,page=4,trim=3cm 4cm 3.5cm 2cm, clip]{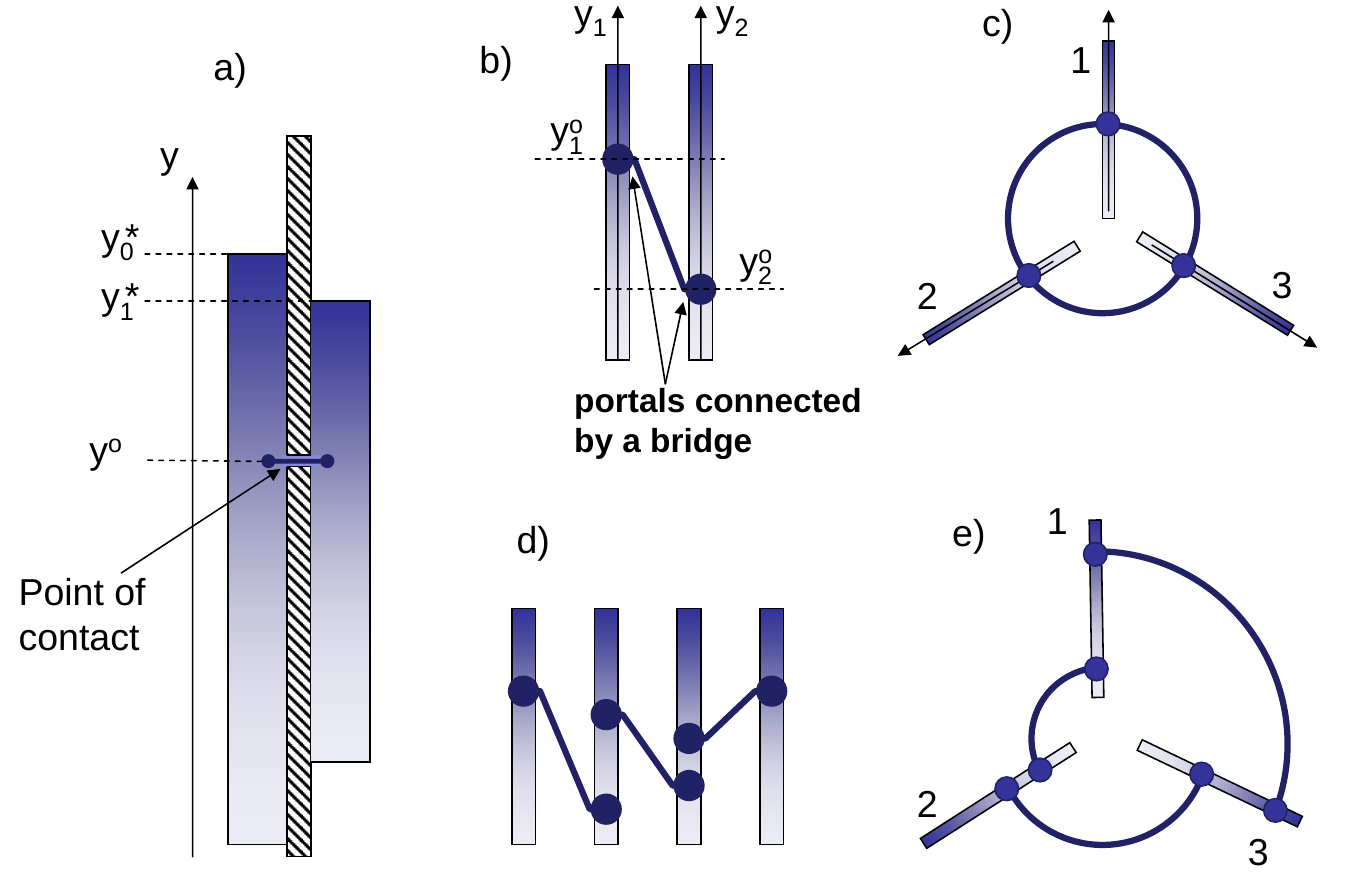}
\caption{Point of contact equilibrium when the subsystems have a) the same ranking and a direct connection,
b) autonomous ranking and agreed connection through portals, c) multiple connections through portals that are unique for each subsystem,
d) multiple connections through multiple portals without forming a loop, and e) multiple connections through multiple 
portals with a loop. Note that the last case can be inconsistent with the point of contact equilibrium. }
\label{fig4}
\end{center}
\end{figure}

Connections through a point of contact can be given different interpretations.
Figure \ref{fig4}a, shows two subsystems with the same property $y$ that are
connected at location $y=y^{\circ}$. Another interpretation, which is
illustrated in figure \ref{fig4}b, is that $y_{1}$ and $y_{2}$ are internal
properties of the subsystems, generally not related to each other, while the
point of contact is an agreement that establishes correspondence of two
locations $y_{1}^{\circ}$ and $y_{2}^{\circ}$ that are called open portals.
Particles can freely move between these portals through the bridge connecting
the portals. Note that subsystems can have more than one open portal (see
figure \ref{fig4}d) as long as connections between these portals do not form a
loop. Figure \ref{fig4}e illustrates such a loop that can make particle
densities at two open portals that belong to a single subsystem inconsistent
with each other. This would change the shapes of particle distributions
$\phi_{I}(\mathbf{y})$ within the subsystems.

The case that is most interesting from the thermodynamic perspective is shown
in figure \ref{fig4}c: each subsystem has only one open portal --- this
ensures that the number of particles $n_{I}$ within each subsystem changes,
while the subsystem distributions $\phi_{I}(\mathbf{y})$ remain the same
(presuming that each subsystem always converges to its internal steady state).
Each portal can be connected to one or more of the portals that belong to the
other subsystems. This connection is characterised by the subsystem particle
numbers $n_{I}$ converging to their equilibrium values and by the detailed
equilibrium between the subsystems (although the detailed balance is not
necessarily achieved for the steady states within each subsystem). Assuming
that the portal $\mathbf{y}_{I}^{\circ}$ of subsystem $I$ is connected to the
portal $\mathbf{y}_{J}^{\circ}$ of subsystem $J$, the equilibrium condition
(\ref{eq-1}) is now rewritten as
\begin{equation}
\frac{n_{I}\phi_{I}(\mathbf{y}_{I}^{\circ})}{A_{I}(\mathbf{y}_{I}^{\circ}%
)}=\frac{n_{J}\phi_{J}(\mathbf{y}_{J}^{\circ})}{A_{J}(\mathbf{y}_{J}^{\circ})}
\label{eq-1-2}%
\end{equation}

Let us consider how this equilibrium between $K$ subsystems can be
characterised by Tsallis entropy, which is defined as
\begin{equation}
S([\varphi])=\sum_{I=1}^{K}\int_{\infty}\left(  \tilde{\varphi}_{I}%
(\mathbf{y})\ln_{q}\left(  \frac{1}{\tilde{\varphi}_{I}(\mathbf{y})}\right)
+\tilde{\varphi}_{I}(\mathbf{y})^{q}s_{I}(\mathbf{y})\right)  A_{I}^{\circ
}(\mathbf{y})d\mathbf{y}\;
\end{equation}
with constraints on the overall number of particles in the system:
\begin{equation}
\sum_{I=1}^{K}n_{I}=n,\;\;\tilde{\varphi}_{I}A_{I}^{\circ}(\mathbf{y}%
)=\varphi_{I}(\mathbf{y})=n_{I}\phi_{I}(\mathbf{y}),\;\int\phi_{I}%
(\mathbf{y})d\mathbf{y}=1,\;
\end{equation}
where $A_{I}^{\circ}(\mathbf{y})$ is the effective phase volume in the
subsystems defined by
\begin{equation}
A_{I}^{\circ}(\mathbf{y})=a_{I}^{\circ}A_{I}(\mathbf{y}) \label{Ao}%
\end{equation}
Here $A_{I}(\mathbf{y})$ is the true phase volume and $a_{I}^{\circ}$ is the
correcting coefficient, which depends on the location of the portal
$\mathbf{y}_{I}^{\circ}.$ The value of $a_{I}^{\circ}$ is determined later
from the equilibrium conditions specified by (\ref{eq-1-2}). Note that the
definition of entropy is extensive with respect to superposition of the
subsystems but is generally non-extensive within each of the subsystems:
\begin{align}
S  &  =\sum_{I=1}^{K}S_{I}(\varphi_{I}),\;\label{SI11}\\
S_{I}  &  =\int_{\infty}\left(  \tilde{\varphi}_{I}(\mathbf{y})\ln_{q}\left(
\frac{1}{\tilde{\varphi}_{I}(\mathbf{y})}\right)  +\tilde{\varphi}%
_{I}(\mathbf{y})^{q}s_{I}(\mathbf{y})\right)  A_{I}^{\circ}(\mathbf{y}%
)d\mathbf{y}\nonumber\\
&  =\int_{\infty}\left(  \tilde{\varphi}_{I}(\mathbf{y})\ln_{q}\left(
\frac{\exp_{q}\left(  s_{I}(\mathbf{y})\right)  }{\tilde{\varphi}%
_{I}(\mathbf{y})}\right)  \right)  A_{I}^{\circ}(\mathbf{y})d\mathbf{y}
\label{SI2}%
\end{align}
As demonstrated by Abe \cite{Abe3}, extensivity of entropy simplifies
equilibrium analyses and removes the need to distinguish nominal and physical
values of intensive properties. The general scaling laws that ensure
extensivity of the entropy have been considered by Hanel and Thurner
\cite{Thurner1}. Here, extensivity with respect to superposition of the
subsystems is simply enforced by the definition of the entropy in (\ref{SI11}).

Maximisation of $S$ is conducted first over for the shape of $\varphi
_{I}(\mathbf{y})$ under constraint
\begin{equation}
\int_{\frak{D}_{I}}\tilde{\varphi}_{I}(\mathbf{y})A_{I}^{\circ}(\mathbf{y}%
)d\mathbf{y}=n_{I} \label{nID}%
\end{equation}
and then with respect to $n_{I}$ under condition $\Sigma_{I}n_{I}=n$. The
first step results in
\begin{equation}
\ln_{q}\left(  \frac{e_{q}^{s_{I}(\mathbf{y})}}{c_{q}\varphi_{I}(\mathbf{y}%
)}\right)  =\lambda_{I}%
\end{equation}
that is
\begin{equation}
\frac{\varphi_{I}(\mathbf{y})}{n_{I}}=\phi_{I}(\mathbf{y})=A_{I}^{\circ
}(\mathbf{y})\frac{\tilde{\varphi}_{I}(\mathbf{y})}{n_{I}}=A_{I}^{\circ
}(\mathbf{y})\frac{e_{q}^{s_{I}(\mathbf{y})}}{c_{q}Z_{I}^{\circ}}%
\end{equation}
where $\lambda_{I}$ are the Lagrangian multipliers associated
with\ constraints (\ref{nID}) and the effective partition function
$Z_{I}^{\circ}$ is determined in terms of the true partition function $Z_{I}$
by the normalisation condition
\begin{equation}
Z_{I}^{\circ}=a_{I}^{\circ}Z_{I},\;\;Z_{I}=\int_{\frak{D}_{I}}\frac
{e_{q}^{s_{I}(\mathbf{y})}}{c_{q}}A_{I}(\mathbf{y})d\mathbf{y}%
\end{equation}
The substitution of $\varphi_{I}(\mathbf{y})$ into (\ref{SI2}) results in the
following problem of finding entropy extremum
\begin{equation}
S=\sum_{I=1}^{K}S_{I}(\varphi_{I}),\;S_{I}=n_{I}\ln_{q}\left(  \frac
{Z_{I}^{\circ}}{n_{I}}\right)  ,\;\;\sum_{I=1}^{K}n_{I}=n,\; \label{Sn}%
\end{equation}
Maximisation of $S$ in (\ref{Sn}) yields
\begin{equation}
\chi_{I}=\frac{\partial S}{\partial n_{I}}=\ln_{q}\left(  \frac{Z_{I}^{\circ}%
}{c_{q}n_{I}}\right)  =\ln_{q}\left(  \frac{a_{I}^{\circ}Z_{I}}{c_{q}n_{I}%
}\right)  =\lambda^{\circ} \label{chi}%
\end{equation}
where $\chi_{I}$ is the competitive potential of $I$-th subsystem and
$\lambda^{\circ}$ is the $I$-independent Lagrangian multiplier associated with
fixing the overall number of particles to $n$ in (\ref{Sn}). The equilibrium
distribution of particles between subsystems is then given by
\begin{equation}
n_{I}=Ca_{I}^{\circ}Z_{I},\;\;C=\frac{\sum_{I}n_{I}}{\sum_{I}Z_{I}^{\circ}}
\label{nI}%
\end{equation}
Note that equations (\ref{chi}) and (\ref{nI})\ imply the detailed equilibrium
$\chi_{I}=\chi_{J}\;\;$for any $I$ and $J.$\ Consistency of (\ref{nI}) with
(\ref{eq-1-2}) determines the correcting coefficients
\begin{equation}
a_{I}^{\circ}=\operatorname{const}\frac{A_{I}(\mathbf{y}_{I}^{\circ})}%
{\phi_{I}(\mathbf{y}_{I}^{\circ})Z_{I}} \label{aIconst}%
\end{equation}
The constant in this equation is arbitrary (since it does not affect the
equilibrium state) and can be set to unity without loss of generality.

Assuming that $A_{I}=1$ and all $Z_{I}$ are the same, we obtain $a_{I}^{\circ
}=1/\phi_{I}(\mathbf{y}_{I}^{\circ})$ and the following expression for the
overall entropy
\begin{equation}
S=\sum_{I=1}^{K}n_{I}\ln_{q}\left(  \frac{1}{n_{I}\phi_{I}(\mathbf{y}%
_{I}^{\circ})}\right)  \;
\end{equation}
Thus, equilibrium through a point of contact results in defining the effective
phase volumes of the subsystems, which are responsible for the equilibrium
conditions. The competitors' perception of the phase volume of a subsystem
depends on location of its portal. The perceived volume is larger and the
subsystem has higher competitive potential $\chi_{I}$ when the portal is
located at lower ranks. Assuming that portal connections are consistent with
true competitiveness of the elements, we conclude that more competitive
subsystems tend to possess higher effective phase volume.

\section{Intransitivity, transition to complexity and the risk/benefit dilemma}

If competition becomes intransitive and intransitive triplets (\ref{int3})
exist, absolute ranking in not possible in such systems and there is no
absolute entropy (since entropy potential is attached to the absolute
ranking). Some intransitive systems may still retain local transitivity in
smaller subdomains. In this case, the system may behave locally in the same
way as transitive systems do and it is still possible to use local absolute
ranking and local entropy. In this case, the analog of the zeroth law of
thermodynamics becomes invalid allowing for intransitivity of competitive
potentials, such as $\chi_{1}\prec\chi_{2}\prec\chi_{3}\prec\chi_{1}$
(consider the subsystems $I=1,2,3$ shown in figure \ref{fig3}b assuming that
these subsystems are connected), and for cyclic evolutions. The system shown
in figure \ref{fig1}b is locally transitive and globally intransitive.
Assuming that some positive mutations are present, this system evolves
transitively in the vicinity of point A by escalating in the direction of
increase of the local ranking, but the overall evolution appears to be cyclic
moving from A to B then from B to C and finally from C back to A . When
intransitivity becomes stronger (denser) and intransitive triplets can be
found in vicinity of any point, even local evolution of the system may become
inconsistent with the principles of competitive thermodynamics. In complex
systems, this evolution may result in competitive degradation (a process
accompanied by slow but noticeable gradual decrease of competitiveness) and in
competitive cooperation (formation of structures with a reduced level of
internal competition and violating the Stosszahlansatz). From the perspective
of competitive thermodynamics, these processes are abnormal (see
refs.\cite{K-PS2012,K-PT2013} for further discussion).

In this section we consider a different example that involves punctuated
evolution: for most of the time the system seems to behave transitively and
escalate towards higher ranks and higher entropy. This escalation is
nevertheless punctuated by occasional crisis events where the state of the
system collapses to (or near to) the ground state. The system then repeats the
slow growth / sudden collapse cycle. Note that only the cyclic component of
evolution is considered here, while competitive evolutions may also involve a
translational component (or components) and become spiral \cite{K-PT2013}.
\emph{Cycles and collapses are common in real-world complex competitive
systems of different kinds \cite{KlimIJTKS2008,Grudin2012}. }

\begin{figure}[h]
\begin{center}
\includegraphics[width=14cm,page=5,trim=3cm 5cm 3.5cm 5cm, clip]{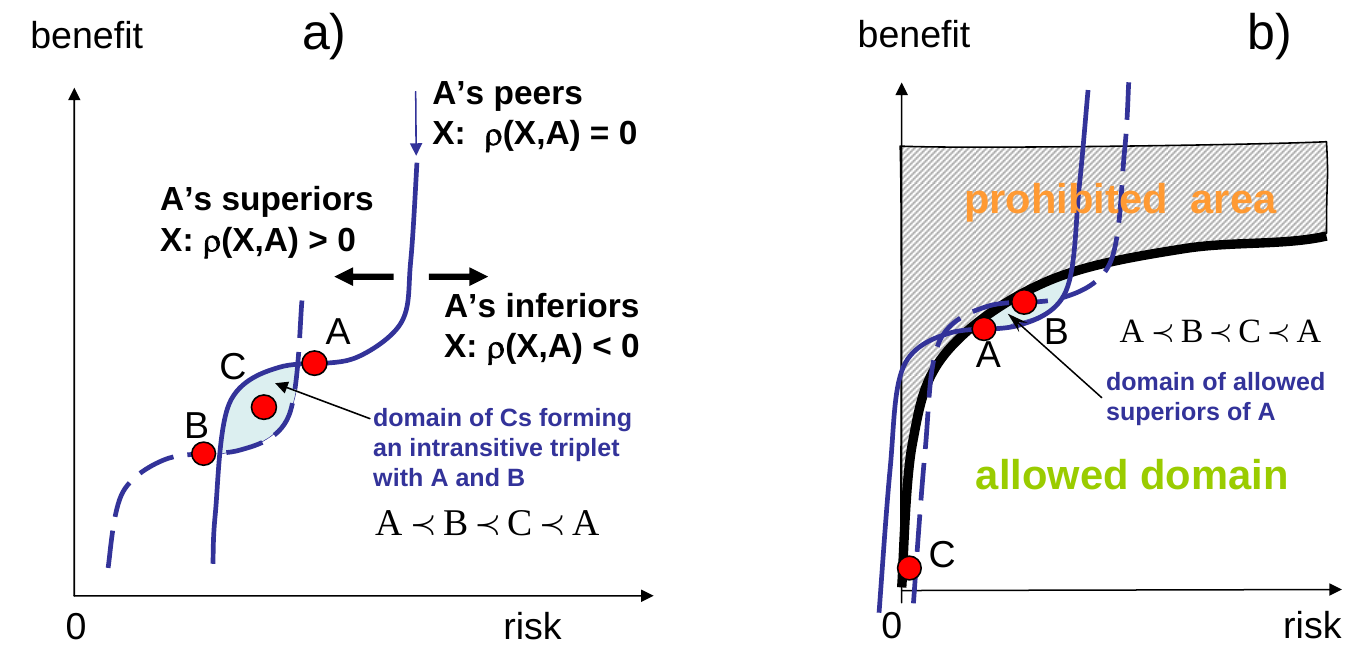}
\caption{Intransitivity in the risk/benefit dilemma: a) strong intransitivity of intransitive triplets A, B and C densely present in the domain; 
b) intransitivity of aggressive strategy B winning over A but losing to defensive strategy C, which is considered to be inferior to A. }
\label{fig5}
\end{center}
\end{figure}

The present example of punctuated evolution is based on the risk/benefit
dilemma (RBD): when comparing the available strategies, we would like to have
low risk and high benefits; hence the problem two parameters: the risk is
denoted by $y^{(1)}$ and the benefit denoted by $y^{(2)}$. While high
$y^{(2)}$ and low $y^{(1)}$ are most attractive, some compromises increasing
risk to increase the benefit or lowering the benefit to lower the risk may be
necessary. When comparing two strategies, $\mathbf{y}_{\text{A}}$ and
$\mathbf{y}_{\text{B}}$, the choice is performed according to the following
coranking
\begin{equation}
\rho\left(  \mathbf{y}_{\text{A}},\mathbf{y}_{\text{B}}\right)  =\left(
y_{\text{A}}^{(2)}-y_{\text{B}}^{(2)}\right)  -\frac{\left(  y_{\text{A}%
}^{(1)}-y_{\text{B}}^{(1)}\right)  ^{m}}{h},
\end{equation}
that is strategy A is preferred over strategy B when $\rho\left(
\mathbf{y}_{\text{A}},\mathbf{y}_{\text{B}}\right)  >0$. We consider two
choices of parameters
\begin{align*}
\text{RBD1}  &  \text{:\ \ }m=3,\;h=1\\
\text{RBD2}  &  \text{:\ \ }m=1,\;h=3
\end{align*}
One can easily see that choice RBD2 is transitive allowing for absolute
ranking
\begin{equation}
\rho\left(  \mathbf{y}_{\text{A}},\mathbf{y}_{\text{B}}\right)  _{\text{RBD2}%
}=r(\mathbf{y}_{\text{A}})-r(\mathbf{y}_{\text{B}}),\;\;\;r(\mathbf{y}%
)=y^{(2)}-y^{(1)}/3 \label{rank}%
\end{equation}
In the case RBD2 our assessment of the risk and the benefit is linear so that
evolutions maximising the absolute ranking $r$ are expected. In case RBD1,
however, we tend to neglect small increases in risk and opt for higher
benefits but a large increase in risk becomes the major concern that
overweights even significant benefits. Choice RBD1 appears to be strongly
(densely) intransitive: as illustrated in figure \ref{fig5}a, there are
intransitive triplets (\ref{int3}) in vicinity of every point. Both cases RBD1
and RBD2 deploy the same mutations, which are predominately small but can be
large on rare occasions. Mutations reaching the prohibited area are banned.

Figure \ref{fig5}b shows the computational domain. The gray area
$y^{(2)}>\left(  y^{(1)}\right)  ^{1/3}$ is prohibited, reflecting the fact
that one cannot have large benefits without being exposed to significant
risks. The strategies superior with respect to A are in the small dark area
causing the system to evolve to higher risks and higher benefits. In
transitive case, the system grows to reach the equilibrium point maximising
the absolute ranking $r(\mathbf{y})$ and then remains in the this state of
relatively high benefits and reasonable risks forever. In intransitive case,
the system does not stay in equilibrium but collapses into a defensive
strategy involving low risks and low benefits. The reason for this collapse is
illustrated in figure \ref{fig5}b. the aggressive strategy A is preferred over
defensive strategy C but as the system evolves even into a more aggressive
strategy B, the risk associated with B becomes too high and at certain moment
defensive strategy C becomes more attractive than B. This results in the
collapse of the growth and rapid transition to defensive strategies.

For the transitive case, the entropy is defined by equation (\ref{Scs}). The
translational case $\gamma=1$ with entropy potential depending linearly on
ranking $s_{y}(\mathbf{y})=kr(\mathbf{y})$ is chosen. The parameters
$q=1/Q=1/1.2$ and $kq=70$ are selected to match the equilibrium distribution
discussed below. The entropy definition takes the following form:
\begin{equation}
S=\int_{\infty}\left(  f(\mathbf{y})\ln_{q}\left(  \frac{1}{f(\mathbf{y}%
)}\right)  +f(\mathbf{y})kr(\mathbf{y})\right)  d\mathbf{y} \label{Sck2}%
\end{equation}
In the case under consideration, the entropy is practically dominated by the
ranking term and the difference between conventional logarithmic entropy and
Tsallis entropy is not large.

Figure \ref{figm4} illustrates intransitive and transitive evolutions in the
risk/benefit dilemma. The transitive branch is obtained by switching
parameters from RBD1 to RBD2 at 410 time steps. The following intransitive and
transitive evolutions seem to be very similar but only up to a point where
maximal $S$ is reached. The same definitions of ranking (\ref{rank}) and
entropy (\ref{Sck2}) are used for both cases, transitive and intransitive.
Then the evolutions diverge: the transitive branch remains in equilibrium
state near the point of maximal entropy and maximal ranking while the
intransitive branch falls down into the region of defensive strategies. Video
files covering these evolutions between steps 1 and 590 is offered as an
electronic supplement to this article (see Appendix for more details).

\begin{figure}[h]
\begin{center}
\includegraphics[width=7.5cm]{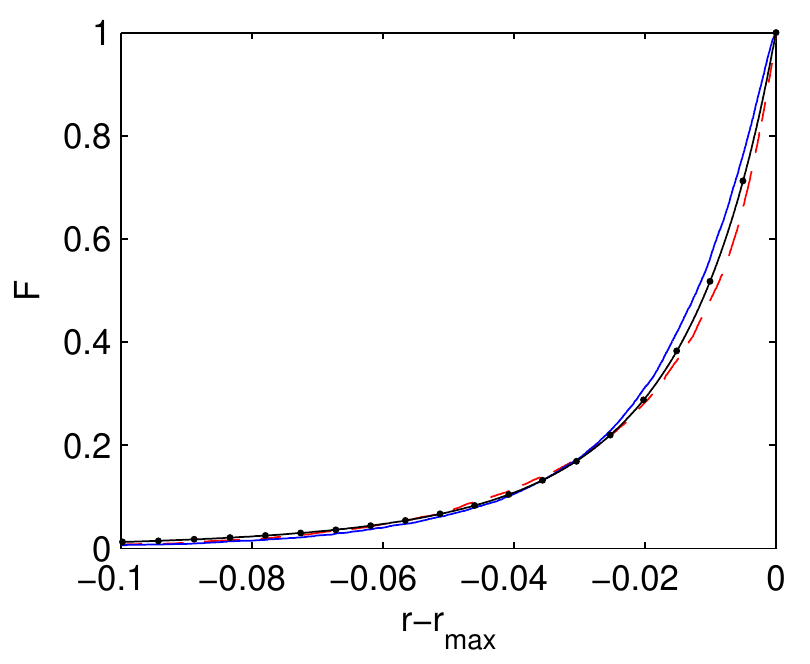}
\caption{Cdf for the rank distribution at 590 time steps (the same simulation as in figure \ref{figm4}, $r_{{\max}} \approx 0.65 $ ).
Solid curve --- intransitive (RBD1) simulation;
dashed curve --- transitive (RBD2) simulation; 
solid curve with dots --- approximation by the $q$-exponent with $Q=1.2$ and $k/Q=70$.}
\label{figm3}
\end{center}
\end{figure}

\begin{figure}[h]
\begin{center}
\includegraphics[width=12cm]{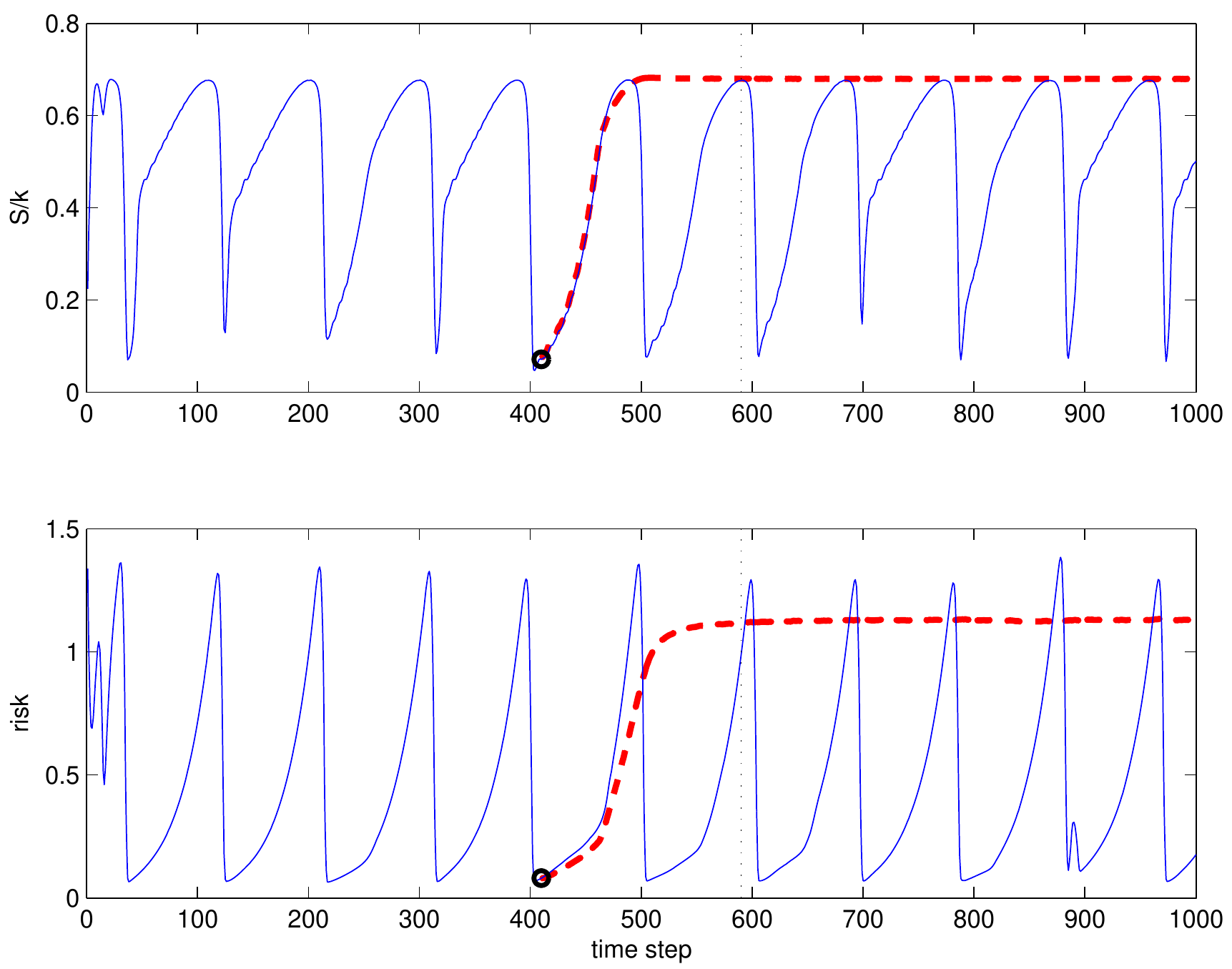}
\caption{Simulations of the risk/benefit dilemma. Solid curve --- intransitive (RBD1) simulation; 
dashed curve --- transitive (RBD2) simulation initiated at 410 time steps. Top figure: normilised entropy $S/k$ versus time steps; 
bottom figure: risk $y^{(1)}$ versus time steps. Vertical dotted line shows 590 time steps. }
\label{figm4}
\end{center}
\end{figure}

If the underlying competition rules and long-term history of the evolution are
unknown, determining how a system is going to behave in the future by
analysing the current distributions may be very difficult. Figure \ref{figm3}
illustrates this point. This figure shows the cdf of ranking $r$ for
transitive evolution (RBD2) and intransitive evolution (RBD1) at 590 time
steps. Both distributions are very similar and can be approximated quite well
by the $q$-exponential cdf (\ref{FQk}) with $Q=1.2$ and $k/Q=70$.

The competitive mechanism represented by the risk/benefit dilemma, can be one
of the forces enacting economic cycles in the real world. From the economic
perspective, the strategies reflected by RBD2 are seen as rational behaviours
of individual players (say, investment agents). The benefit is represented by
returns on investments and ranking $r$ is conventionally called
\textit{utility} in economics \cite{Mehta2001}. This utility weights different
factors against each other and enforces transitivity of economic decisions.
The intransitive strategies reflected by RBD1 would be seen by economists as
semi-rational. Since risk and benefit do not represent directly comparable
categories and evaluation of risk is always subject to greater uncertainty,
overlooking small risks and being overly concerned with high risks is a
plausible economic strategy for any individual or company. While switching
from the linear RBD2 to non-linear RBD1 seems like a minor adjustment for an
economic element, it has a major effect on functioning of the whole system:
economic growth is interrupted by collapses and the system evolves cyclically.
Competition forces the competing elements to take higher and higher risks
until the risk becomes unsustainable.

\section{Conclusions}

In competitive systems with Gibbs mutations, the distributions tend to be
exponential (assuming isotropy of the property space). This case is described
by the strongest similarity to conventional thermodynamics and the existence
of detailed balance in the system. When the distribution of mutations deviates
from that of Gibbs mutations, the $q$-exponents become very good
approximations characterising the existence of long or short tails in the
distributions caused by biases in taxing and supplying. In competitive
thermodynamics, this corresponds to replacing conventional Boltzmann-Gibbs
entropy by Tsallis entropy.

Unlike in conventional thermodynamics, competitive systems allow different
types of equilibria possessing different degrees of similarity with the
conventional thermodynamic equilibrium. Competition between subsystems without
exchange of mutations tends to be less stable than the connected equilibria
where subsystems exchange particles through both competition and mutations.
Among connected equilibria, the case of Gibbs mutations bears the highest
resemblance to conventional thermodynamics. When mutations are not of the
Gibbs type, the point of contact equilibrium preserves this resemblance more
than the other cases. The point of contact equilibrium has been analysed using
Tsallis entropy. This analysis results in equilibrium conditions determined by
equivalence of competitive potentials. These potentials are linked to the
introduced effective phase volumes of subsystems that depend on location of
the point of contact.

The thermodynamic analogy requires transitivity of competition rules. In case
of intransitive competition rules, the system may behave anomalously when
considered from the perspective of competitive thermodynamics. This involves
formation of structures, competitive degradations and cycles. The present work
uses the example of the competitive risk/benefits dilemma and analyses the
case of punctuated evolutions. For most of the time, the evolution of an
intransitive competitive system, which represents the dilemma, closely
resembles evolutions of transitive systems, which increase ranking and the
associated entropy. At some moments, however, this evolution is punctuated and
results in an abrupt collapse, which decreases ranking and the associated
competitive entropy --- this cannot possibly happen when the competition is
transitive. Then the system starts to grow and repeats the cycle again. While
consideration of competitive processes in this work is generic, similar
behaviours can be found in biological, economic and other systems.

\bigskip

\acknowledgements{Acknowledgements} \ The author thanks Bruce Littleboy for
insightful discussions of economic issues. The author acknowledges funding by
the Australian Research Council.

\bibliographystyle{mdpi}
\bibliography{comp}

\appendix

\section*{APPENDIX: video files with the simulations of the risk/benefit dilemma.}

Simulations of the cases RBD1 and RBD2 involving 10000 Pope particles are
offered as video supplements to the journal version of this article: 

\begin{itemize}
\item  RBD1.avi, 3MB, 1-590 steps,

\item  RBD2.avi, 1MB, 410-590 steps.
\end{itemize}

\begin{figure}[h]
\begin{center}
\includegraphics[width=14cm,page=6,trim=2cm 2cm 4cm 5cm, clip]{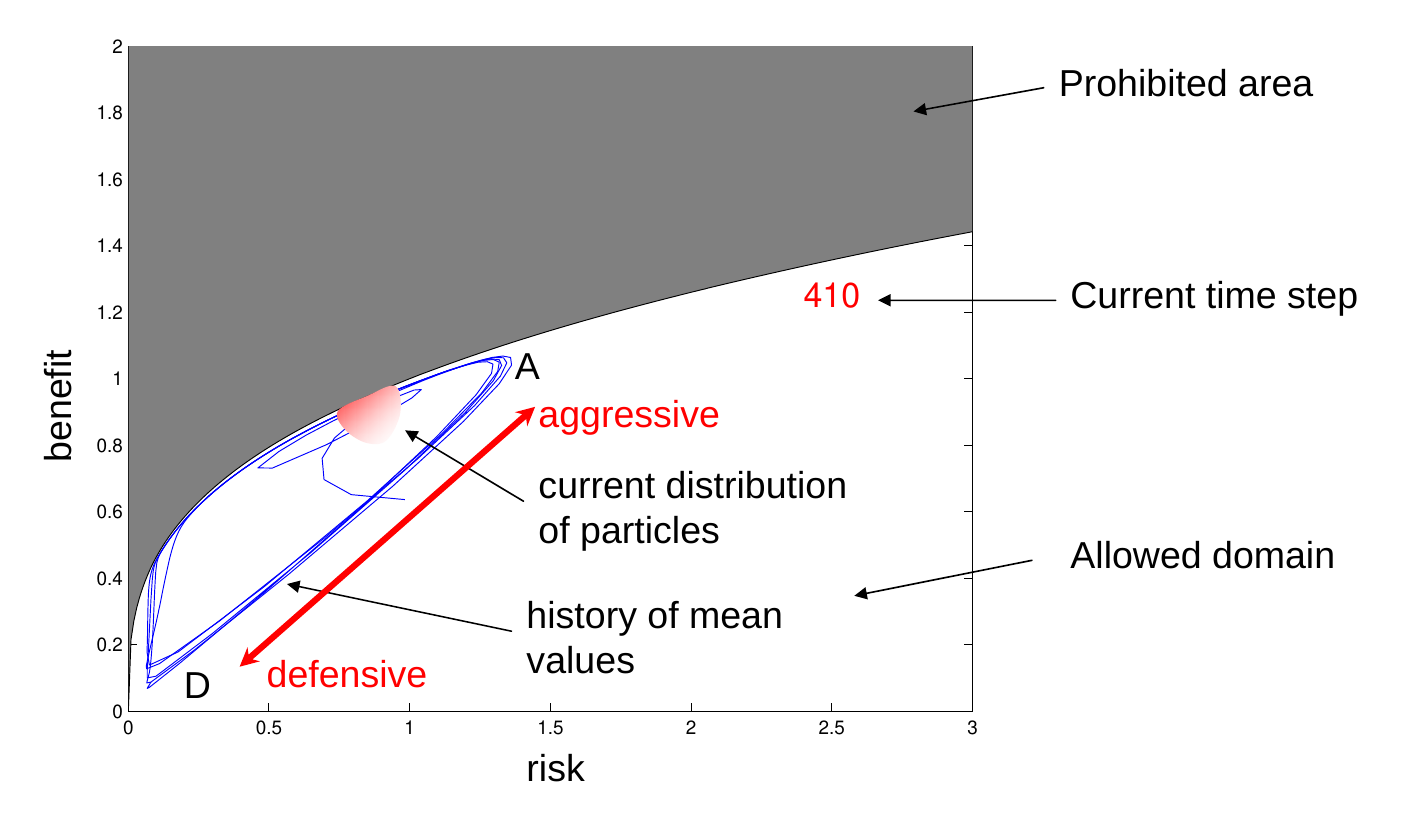}
\caption{Simulations of the risk/benefit dilemma: notations used in the video files.}
\label{figA}
\end{center}
\end{figure}

Competition is intransitive in RBD1 and transitive in RBD2. The cases are
branched apart at 410 time steps with the same distribution of particles. The
format of the videos is explained in figure \ref{figA}. In the intransitive
simulations of the risk/benefits dilemma, competition forces competitors to
undertake more and more aggressive strategies, while the distribution moves
from D to A. This leads to unsustainably high risk and punctuation of
continuous evolution by a sudden collapse of the system by elements seeking
refuge in defensive strategies near D. While the evolution is punctuated in
the intransitive case, the transitive version of the simulations safely
reaches equilibrium and remains there forever. In spite of principal
differences, the ascending fragments of both simulations are very similar.

\pagebreak 

\end{document}